\documentclass{wileyarticle}

\articletype{STARS2017/SMFNS2017 Conference Proceedings}%
\artid{}%
\jid{}%
\jvol{}
\jyear{}%
\jissue{}%
\jmonth{}%
\doi{}%
\copyeditor{}%
\startpage{1}%
\received{}
\revised{}
\accepted{}

\def\K{{\rm K}}
\def\cm{{\rm cm}}

\def\sec{{\rm s}}

\def\g{{\rm g}}
\def\G{{\rm G}}
\def\yr{{\rm yr}}
\def\fm{{\rm fm}}

\def\rp{{\rm p}}
\def\ft{{\rm ft}}

\begin{document}

\title{Fluxtube Dynamics in Neutron Star Cores}

\author[1]{Vanessa Graber}

\authormark{Graber}

\address[1]{\orgdiv{Department of Physics and McGill Space Institute}, \orgname{McGill University}, \orgaddress{\state{Montreal}, \country{Canada}}}

\corres{Department of Physics, McGill University, 3600 rue University, Montreal, QC, H3A 2T8, Canada \email{vanessa.graber@mcgill.ca}}

\begin{abstract}%
Although the detailed structure of neutron stars remains unknown, their equilibrium temperatures lie well below the Fermi temperature
of dense nuclear matter, suggesting that the nucleons in the stars' core form Cooper pairs and exhibit macroscopic quantum behavior.
The presence of such condensates impacts on the neutron stars' large scale properties. Specifically, superconducting protons in the outer
core (expected to show type-II properties) alter the stars' magnetism as the magnetic field is no longer locked to the charged plasma
but instead confined to fluxtubes. The motion of these structures governs the dynamics of the core magnetic field. To examine if field
evolution could be driven on observable timescales, several mechanisms affecting the fluxtube distribution are addressed and characteristic
timescales for realistic equations of state estimated. The results suggest that the corresponding timescales are not constant but vary
for different densities inside the star, generally being shortest close to the crust-core interface.
\end{abstract}


\maketitle


\section{Introduction}
\label{sec-Intro}

Neutron stars host some of the strongest magnetic fields in the Universe. Field strengths (inferred from the stars' dipole spin-down) reach
up to $10^{15} \, \G$ for slowly rotating magnetars, whereas typical radio pulsars have $10^{10}-10^{13} \, \G$ and recycled millisecond
pulsars exhibit $10^{8}-10^{10} \, \G$. Such fields are expected to strongly influence the dynamics and could provide a natural explanation
for various observational features. Long-term field evolution could, for example, be responsible for field changes in regular pulsars (observed
to occur on the order of $10^7 \, \yr$~\citep{Lyne1985, Narayan1990}), while magnetic field decay on a timescale of about $10^4 \, \yr$ is
generally considered as the driving force behind the high activity of magnetars~\citep{Thompson1995}. Additionally, an evolving
magnetic field could connect the different classes of neutron stars and provide evolutionary links between them~\citep{Kaspi2010, Harding2013}.
Despite this, field evolution processes are poorly understood. It also remains unclear if the core, which retains a large fraction
of the star's magnetic energy but is often ignored in theoretical studies~\citep{Pons2007, Vigano2013, Gourgouliatos2014}, sources the currents
that generate the magnetic field and thus takes part in the evolution. If this is indeed the case, there are indications that the core's
high conductivity essentially brings field evolution to a halt~\citep{Graber2015, Elfritz2016}, leaving the question of how one can reconcile
observed magnetic field changes with the theoretical models.

\begin{figure}[t]
\SPIFIG{\includegraphics[width=230pt]{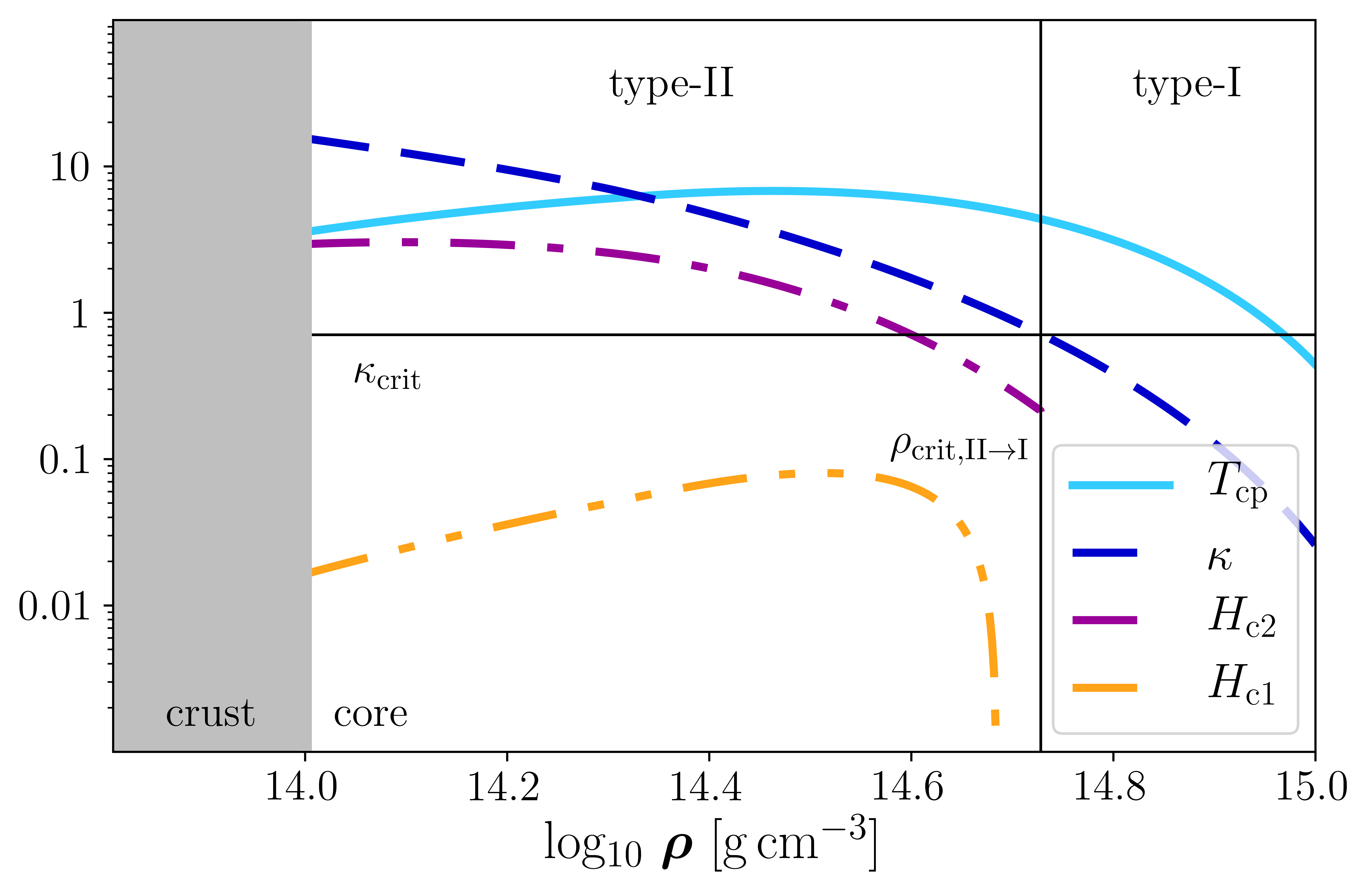}}{
\caption{Parameters of neutron star superconductivity. Shown are $T_{\rm cp}$ (cyan, solid) (normalized to $10^{9} \, \K$), $\kappa$ (blue,
dashed) and $H_{\rm c2}$ (purple, dot-dashed) and $H_{\rm c1}$ (yellow, dot-dot-dashed) (normalized to $10^{16} \, \G$). The horizontal and
vertical line mark $\kappa_{\rm crit}$ and $\rho_{\rm crit,II \to I}$. The cross-section is given for the NRAPR effective EoS~\citep{Steiner2005}
and the energy gap parametrization of~\citet{Ho2012b}. \label{fig1}}}
\end{figure}


\section{Type-II superconductivity}
\label{sec-Supercon}

Equilibrium neutron stars have $10^6 - 10^8 \, \K$, while the nucleon Fermi temperature is $T_{\rm F} \sim 10^{12} \, \K$, implying that these
objects are cold enough to contain superfluid neutrons and superconducting protons. The formation of such quantum condensates can be understood
within the standard theory of laboratory superconductors~\citep{Bardeen1957}. More precisely, the fermions form Cooper pairs due to an attractive
contribution to the nucleon-nucleon interaction and detailed calculations give proton transition temperatures in the range of $T_{\rm cp} \sim
10^9-10^{10} \, \K$~\citep{Ho2012b}. The type of superconductivity depends on the characteristic length-scales involved. Estimating the
Ginzburg-Landau parameter (the ratio of the penetration depth $\lambda$ and the coherence length $\xi_{\rm ft}$) leads to
\begin{equation}
  \kappa = \frac{\lambda}{\xi_{\rm ft}} \approx 3.3 \left( \frac{m_\rp^*}{m} \right)^{3/2}
    \rho_{14}^{-5/6} \left(\frac{x_\rp}{0.05} \right)^{-5/6} \left( \frac{T_{\rm cp}}
    {10^9 \, \K}\right).
\end{equation}
This is larger than the critical value $\kappa_{\rm crit} \equiv 1/\sqrt{2}$ and suggests that the neutron star interior is in a type-II state.
Here $m$ denotes the baryon mass, $\rho_{14} \equiv \rho/(10^{14} \, \g \, \cm^{-3})$ the normalized total mass density and $x_\rp$ the proton
fraction. Moreover, $m_\rp^*$ is the proton effective mass, differing from the bare mass $m$ due to entrainment, a non-dissipative interaction
present in strongly coupled Fermi systems~\citep{Andreev1976}. The corresponding critical fields read
\begin{align}
	H_{\rm c1} & = \frac{\phi_0}{4 \pi \lambda^2} \, \text{ln} \kappa \approx 1.9 \times 10^{14} \left( \frac{m}{m_\rp^*} \right) \rho_{14}
		\left( \frac{x_{\rm p}}{0.05} \right) \, \G, \\[1.2ex]
	H_{\rm c2} &= \frac{\phi_0}{2 \pi \xi_{\rm ft}^2} \approx 2.1 \times 10^{15} \left( \frac{m_\rp^*}{m} \right)^{2} \rho_{14}^{-2/3}
    \nonumber \\&\times
		\left( \frac{x_{\rm p}}{0.05} \right)^{-2/3}  \left( \frac{T_{\rm cp}} {10^9 \, \K} \right)^{2} \, \G,
\end{align}
where $\text{ln} \kappa \simeq 2$~\citep{Tinkham2004} is used to estimate $H_{\rm c1}$. The density-dependent behavior of these parameters is
illustrated in Fig.~\ref{fig1} for the NRAPR effective equation of state~\citep{Steiner2005}. Note that most pulsars have $B \lesssim
H_{\rm c1}$ and would in principle want to expel magnetic flux from their interior. However as argued in the seminal paper of~\citet{Baym1969},
the conductivity of normal matter is so large that the superconducting transition has to occur at constant flux, implying that the outer core
is in a meta-stable type-II state and penetrated by fluxtubes. Fig.~\ref{fig1} further shows that at some density $\rho_{\rm crit,II \to I}$,
$\kappa$ falls below $\kappa_{\rm crit}$ and an intermediate type-I state is present. As these macroscopic regions of zero and non-zero magnetic
flux would be irregularly distributed, it is not obvious how to model such systems theoretically. We thus focus on the outer type-II region,
where quantized fluxtubes are arranged in a hexagonal array. Since each fluxtube carries a unit of flux, $\phi_0 \approx 2 \times 10^{-7} \,\G
\, \cm^2$, the macroscopic magnetic induction $B$ in the star's core is simply obtained by summing all individual flux quanta. This allows one
to relate $B$ to the fluxtube surface density and inter-fluxtube distance:
\begin{align}
  \mathcal{N}_\ft &= \frac{B}{\phi_0} \approx 4.8 \times 10^{18} \, B_{12} \, \cm^{\!-2}, \\
  d_\ft &\simeq \mathcal{N}_\ft^{-1/2} \approx 4.6 \times 10^{-10} \, B_{12}^{-1/2} \, \cm,
\end{align}
where $B_{12} \equiv B/(10^{12} \, \G)$.
This shows that core field evolution is closely linked to the distribution of fluxtubes and mechanisms driving these structures out of the core
towards the crust (where flux subsequently decays) could provide the means to decrease the field strength in the interior. Several effects that
influence the fluxtube behavior are analyzed in the following and studied for realistic equations of state (EoSs).


\section{Fluxtube Coupling Mechanisms}
\label{sec-FTMechanisms}

\subsection{Resistive drag}
\label{subsec-Resistivity}

\begin{figure}[t]
\SPIFIG{\includegraphics[width=230pt]{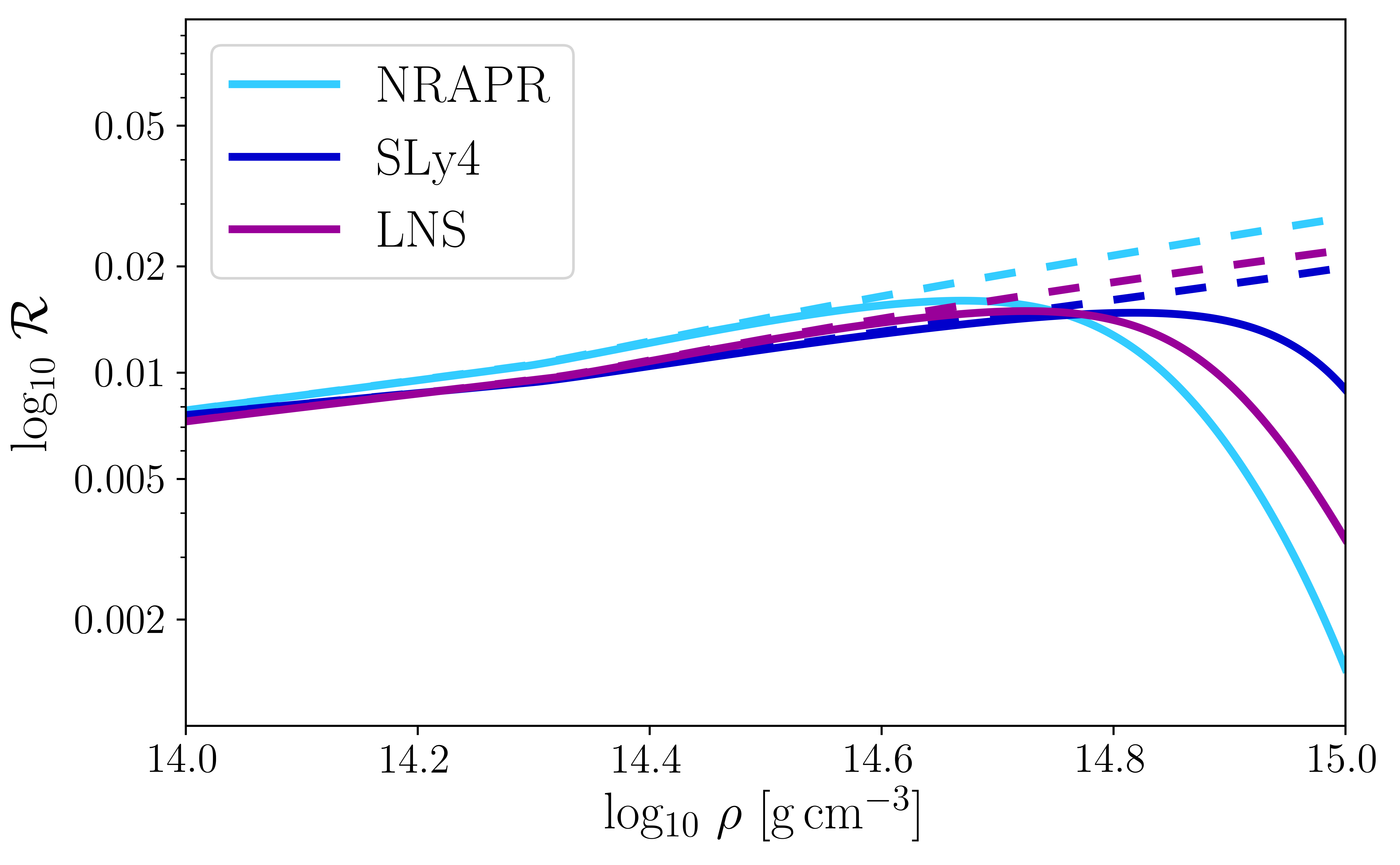}}{
\caption{Behavior of the resistive drag coefficient as a function of density calculated for three different EoSs and singlet proton pairing. The
dashed lines correspond to an approximate solution often found in the literature, which is independent of the energy gap and given by
$\mathcal{R} \approx  3 \pi^2 / (64 \lambda k_{\rm Fe})  \approx 7.9 \times 10^{-3} (m/ m_\rp^*)^{1/2} \rho_{14}^{1/6} ( x_\rp/0.05)^{1/6} $.
\label{fig2}}}
\end{figure}

Electrons can scatter off the fluxtubes' magnetic field~\citep{Alpar1984}, often referred to as mutual friction in analogy with superfluid
hydrodynamics. On mesoscopic scales this drag is proportional to the relative velocity of both components and fully determined by a coefficient
$\mathcal{R}$; giving $\mathbf{f}_{\rm d} = \rho_\rp \kappa \mathcal{R} (\mathbf{v}_{\rm e} - \mathbf{v}_{\rm ft} )$. Here $\rho_\rp$ is the
proton mass density and $\kappa \approx 2.0 \times 10^{-3} \, \cm^{2} \, \sec^{-1}$ the quantum of circulation. Using the formalism
of~\citet{Sauls1982} $\mathcal{R}$ is found to be
\begin{align}
 \mathcal{R} &= \frac{1}{\mathcal{N}_{\rm ft} \kappa} \, \frac{E_{\rm Fe}}{m c^2} \, \frac{1}{\tau} \nonumber \\
 &\approx 1.6 \times 10^{-2} B_{12}^{-1}  \left( \frac{k_{\rm Fe}}{0.75 \, \fm^{-1}}\right)
   \left( \frac{10^{-15} \, \sec}{\tau} \right),
\end{align}
where $k_{\rm Fe}$ denotes the Fermi wave number and $\tau$ the velocity relaxation timescale of the electrons. Given an EoS and superconducting
gap, $\tau$ and $\mathcal{R}$ can be calculated as a function of the star's density. Results for three different parametrized EoSs, namely NRAPR,
SLy4 and LNS (see~\citet{Chamel2008} for details), together with the proton gap parametrization of~\citet{Ho2012b} are shown in Fig~\ref{fig2}. In
the outer core region, the three EoSs do not differ significantly. Note that it has been shown in~\citet{Graber2015} by deriving an
induction equation for type-II superconducting matter that this resistive drag on its own is too weak to result in field decay on short timescales.

\subsection{Repulsive interaction}
\label{subsec-Repulsion}

\begin{figure}[t]
\SPIFIG{\includegraphics[width=230pt]{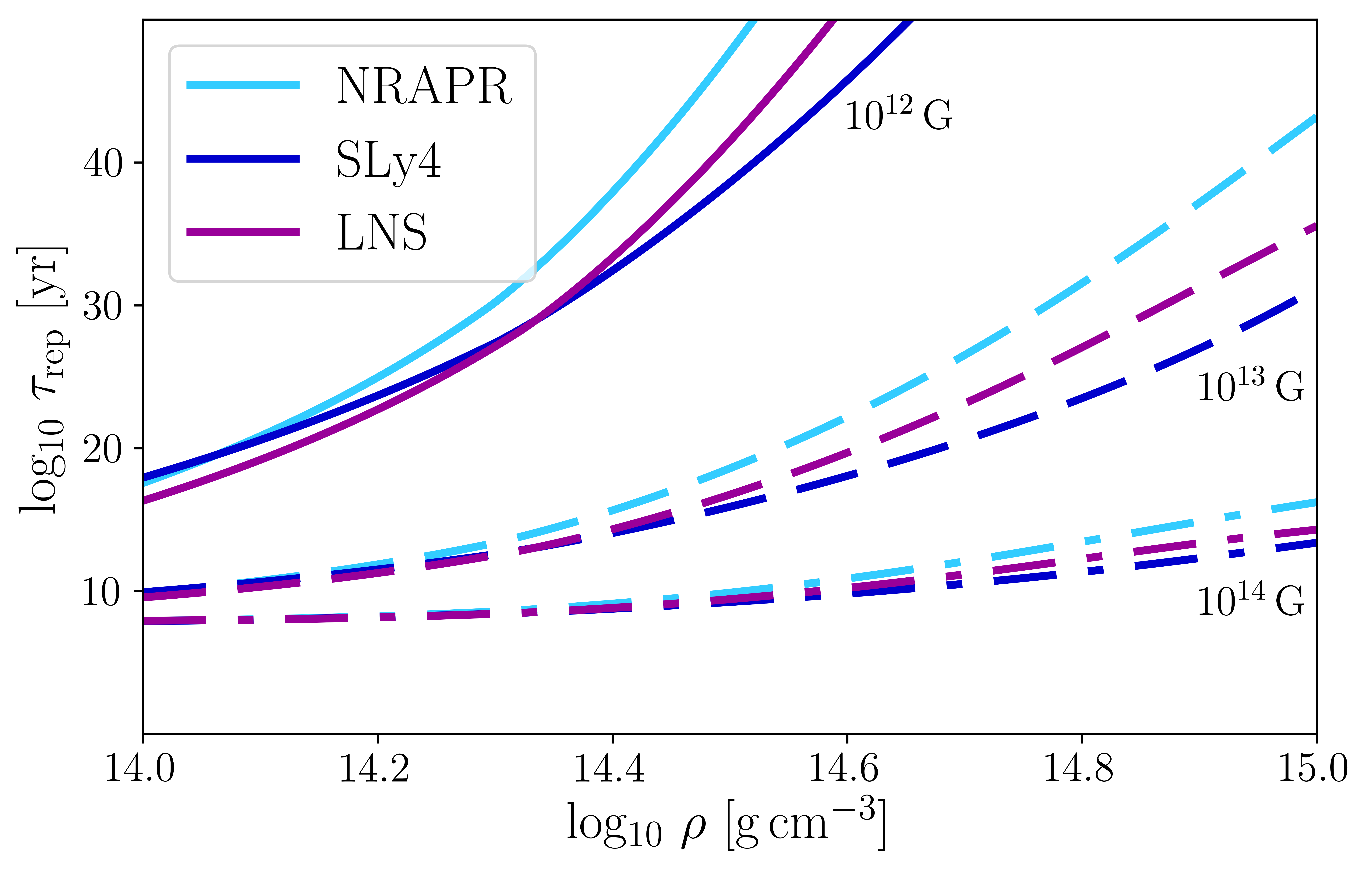}}{
\caption{Density-dependence of the diffusion timescale estimated from balancing the resistive and the repulsive force. Results are given
for three EoSs, three initial field strengths and $L=10^6 \, \cm$. \label{fig3}}}
\end{figure}

The fluxtubes can also be affected by a repulsive interaction between individual lines. Ignoring their detailed structure, one can determine the
interaction energy $\mathcal{E}_{\rm int}$ of two parallel fluxtubes separated by a distance $r_{21}$ and subsequently obtain the  standard
result for the repulsive force acting on a unit length of fluxtube 1 due to the presence of fluxtube 2~\citep{Tinkham2004}
\begin{equation}
 	\mathbf{f}_{\rm rep} = - \nabla \mathcal{E}_{\rm int} = - \frac{\phi_0^2}{8 \pi^2 \lambda^3} \, K_1 \left( \frac{r_{21}}{\lambda} \right)
  \hat{\boldsymbol{r}}_{21}.
\end{equation}
Here, $K_1$ is a modified Bessel function of second kind and $\hat{\boldsymbol{r}}_{21}$ the unit vector pointing from fluxtube 1 to 2. To
generalize this repulsion to a fluxtube lattice, individual contributions simply have to be summed up. For a perfectly hexagonal array, the
net force on each line would exactly vanish and no field changes take place. However, due to the large number of fluxtubes it is likely that
some irregularity affects the type-II state in neutron stars; in analogy with laboratory systems we would specifically expect the long-range
order of the lattice to be destroyed. This would cause a gradient in the fluxtube density, directly related to a non-zero net force on the
fluxtubes and would thus drive field evolution. In an averaged picture, the resulting force should be of the form $\mathbf{f}_{\rm rep} = -
g(\mathcal{N}_{\rm ft}) \nabla \mathcal{N}_{\rm ft}$. For pulsars with $B \lesssim 10^{14} \, \G$ the calculation of $g(\mathcal{N}_{\rm
ft})$ simplifies due to the hierarchy of the relevant length-scales. Since $d_{\rm ft} \simeq r_{21} \gtrsim \lambda$, $K_1$ can be approximated
as a decaying exponential and the summation reduced to the six nearest neighbors. Looking at the detailed geometry, one arrives at
\footnote{We note that the expression~\eqref{eqn-gfunction} does not agree with the equivalent result given by~\citet{Kocharovsky1996}. We further point
out that~\citet{Istomin2016} who study field evolution in accreting neutron stars are likely overestimating the effect of the repulsive force
due to an erroneous approximation of $\mathbf{f}_{\rm rep}$ in the limit $d_{\rm ft} \ll \lambda$.}
\begin{equation}
g (\mathcal{N}_{\rm ft})  \simeq \frac{3 \phi_0^2 }{32 \sqrt{2} \pi^{3/2}}
 \left( \frac{\sqrt{2} \, \mathcal{N}_{\rm ft}^{-1/2}}{3^{1/4} \lambda} \right)^{7/2} \hspace{-0.2cm}
 e^{- \frac{\sqrt{2} \, \mathcal{N}_{\rm ft}^{-1/2}}{3^{1/4} \lambda}}.
 \label{eqn-gfunction}
\end{equation}

We now consider a simple scenario, where the resistive and repulsive mechanisms affect the type-II state. Neglecting fluxtube inertia, the
force balance in this steady state reads $\sum \mathbf{f} = - g(\mathcal{N}_{\rm ft}) \nabla \mathcal{N}_{\rm ft} - \rho_\rp \kappa \mathcal{R}
\mathbf{v}_{\rm ft} = 0$. Solving this for $\mathbf{v}_{\rm ft}$, the resulting expression can be combined with the continuity equation for
$\mathcal{N}_{\rm ft}$ (see~\citet{Glampedakis2011a} for details) to obtain a non-linear diffusion equation for the evolution of the fluxtubes,
i.e. the magnetic induction:
\begin{equation}
  0 = \partial_t \mathcal{N}_{\rm ft} - \nabla \left( \frac{\mathcal{N}_{\rm ft} \, g(\mathcal{N}_{\rm ft})}{\rho_\rp \kappa \mathcal{R}}
   \, \nabla \mathcal{N}_{\rm ft} \right).
   \label{eqn-diff}
\end{equation}
Instead of solving this equation, we simply extract a diffusion timescale. For a characteristic length-scale $L$, one has
 \begin{equation}
 	\tau_{\rm rep} = \frac{L^2 \rho_\rp \kappa \mathcal{R}} {\mathcal{N}_{\rm ft} \, g(\mathcal{N}_{\rm ft})}.
 \end{equation}
Estimates for different EoSs, $L = 10^6 \, \cm$ and three initial magnetic inductions are illustrated in Fig.~\ref{fig3}, showing strong
variability with density. For all field strengths, $\tau_{\rm rep} $ is smallest at low densities and increases significantly towards higher
densities as a result of the exponential in Eqn.~\eqref{eqn-gfunction}. For standard pulsars, $\tau_{\rm rep}$ exceeds the age of the
Universe and no field evolution takes place, but it decreases for higher fields and reaches $\sim 10^7 \, \yr$ for $10^{14} \, \G$. Taking
into account that this estimate sensitively depends on $L$, which could also take values of $10^5 \, \cm$~\citep{Lander2013},
Eqn.~\eqref{eqn-diff} could potentially capture field evolution in magnetars. Note that Eqn.~\eqref{eqn-gfunction} is however not suitable to model
fields $B \gtrsim 10^{14} \, \G$, since then $r_{21} \lesssim \lambda$ and the assumptions in deriving $g(\mathcal{N}_{\rm ft})$ are no longer
valid. Studying this regime will be left for future work. One final note of caution: the repulsive force does not necessarily have to expel
the field out of the star's core, i.e. lead to field decay. Instead this mechanism results in fluxtube motion in the direction opposite to
the density gradient $\nabla \mathcal{N}_{\rm ft}$ and thus strongly depends on the (unknown) details of the magnetic field configuration.


\subsection{Buoyancy}
\label{subsec-Buoyancy}

\begin{figure}[t]
\SPIFIG{\includegraphics[width=230pt]{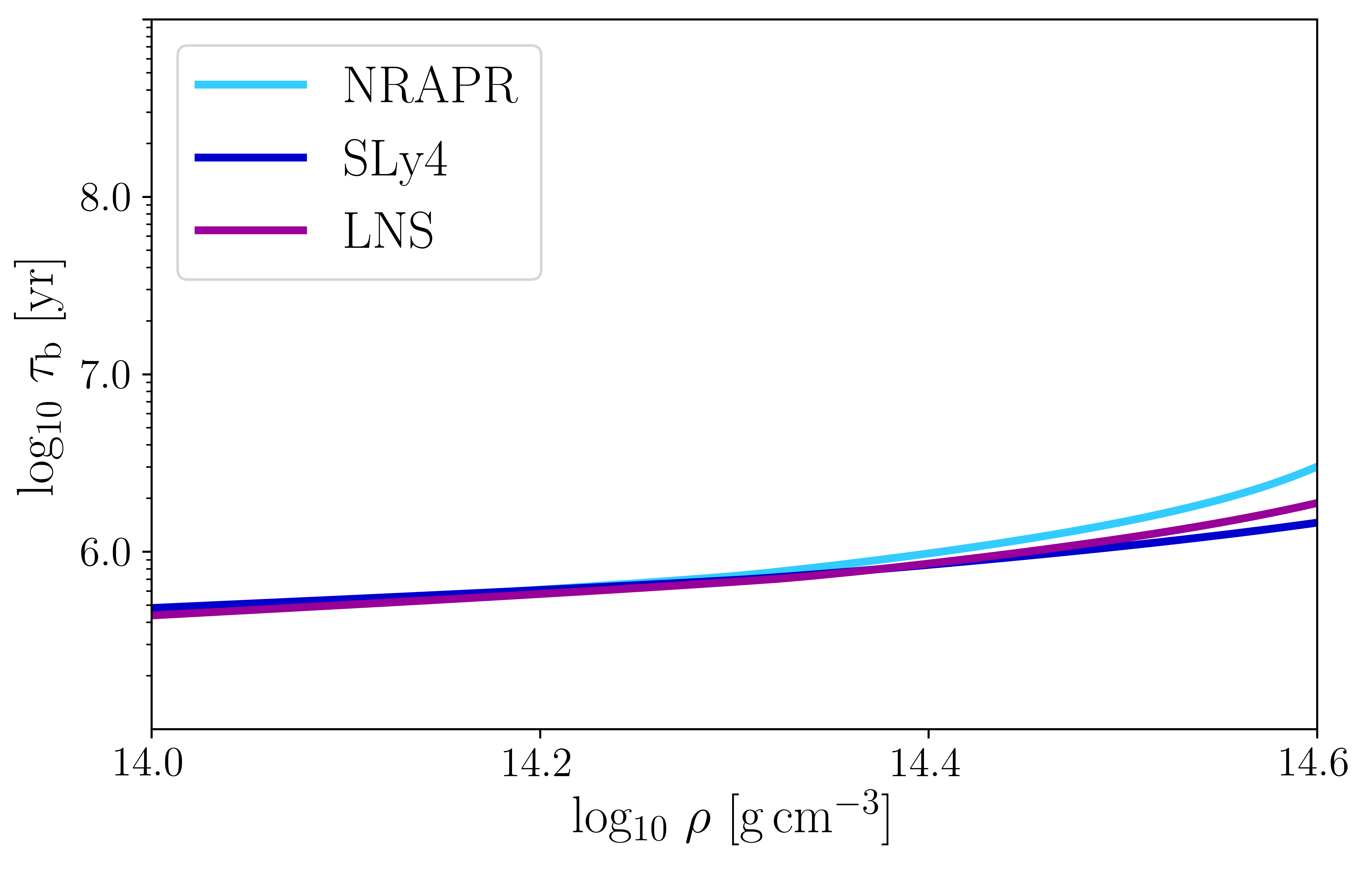}}{
\caption{Density-dependence of the diffusion timescale estimated from balancing the resistive and the buoyancy force. Results are given
for three EoSs, three initial fields and $L=10^6 \, \cm$. \label{fig4}}}
\end{figure}

Fluxtubes are buoyant structures as a result of the magnetic pressure inside their cores. This creates a radially acting lift force, $f_{\rm b}$,
trying to drive the fluxtubes out of the core~\citep{Muslimov1985, Harvey1986}.\footnote{See also~\citet{Dommes2017} for a discussion of the buoyancy
force in a two-component system.} The buoyancy force can be related to the gradient of the superconducting magnetic pressure, which in the limit
$B \lesssim H_{\rm c1}$ satisfied by the neutron star is given by $P = H_{\rm c1} B/4 \pi$~\citep{Easson1977}. Per unit length of fluxtube, one finds
\begin{equation}
  f_{\rm b} = \frac{|- \nabla P|}{\mathcal{N}_{\rm ft}} \simeq  \frac{H_{\rm c1} B}{ \mathcal{N}_{\rm ft} 4\pi L} = \frac{H_{\rm c1} \phi_0}{4 \pi L}.
\end{equation}
Balancing the resistive drag with the buoyancy force, one can arrive at an analogous non-linear diffusion equation as for the repulsive interaction.
The respective timescale now reads
\begin{equation}
  \tau_{\rm b} = L^2 \rho_\rp \kappa \mathcal{R} \, \frac{16 \pi^2 \lambda^2}{\phi_0^2 \text{ln} \kappa }.
\end{equation}
Estimates for $L = 10^6 \, \cm$ and three EoSs are given in Fig.~\ref{fig4}. This shows that $\tau_{\rm b}$ is shortest close to the crust-core
interface with a minimum of $\sim 10^5 \, \yr$ and increases by about one order of magnitude towards the inner core. These timescales are of the order
of observed field changes and buoyancy could potentially explain the physics behind the field evolution. However note that recent self-consistent
magneto-thermal simulations of superconducting cores by~\citet{Elfritz2016} indicate that buoyancy is too weak to drive observable field evolution.


\section{Discussion and conclusions}
\label{sec-Conclusions}

We have studied various mechanisms expected to affect superconducting fluxtubes in the outer neutron star core and calculated characteristic
timescales for these processes. Using three realistic EoSs these timescales were estimated for the cross-section of a neutron star and the
resulting magnetic field changes were found to act on shortest timescales at low densities, close to the crust-core interface. While our simple
estimates suggest that repulsive interaction and buoyancy might be able to drive field evolution on observable timescales, it is difficult to
translate this to realistic systems since more work is needed to account for additional physics that likely affect the neutron star's magnetism.
This specifically involves the question of how fluxtubes interact with neutron vortices that result from the quantization of the neutron
superfluid's rotation in the outer core. Further, it remains unclear if the outer core is indeed in a pure type-II or a mixed type-II/type-I
state~\citep{Link2003, Charbonneau2007, Alford2008}. The presence of such a regime where macroscopic flux-free regions alternate with type-II
domains should strongly depend on the star's initial magnetic flux distribution and the microphysics of the superconducting phase transition. Exploring
the analogy with laboratory condensates could be beneficial in answering this questions~\citep{Graber2017}.


\begin{bm}[Acknowledgments]
VG is supported by a McGill Space Institute postdoc fellowship and the Trottier Chair in Astrophysics
and Cosmology.
\end{bm}

\section*{References}

\bibliography{library_SMFNS2017}

\end{document}